\documentclass[twocolumn,abs,prb,showpacs]{revtex4-1}

\usepackage{graphicx}
\usepackage{dcolumn}
\usepackage{float}
\usepackage{amsmath}
\usepackage{bm}

\newcommand{\comment}[1]{}

\begin{document}


\title{Magneto-Optical Conductivity of Silicene and Other Buckled Honeycomb Lattices}

\author{C.J. Tabert}
\author{E.J. Nicol}
\affiliation{Department of Physics, University of Guelph,
Guelph, Ontario N1G 2W1 Canada} 
\affiliation{Guelph-Waterloo Physics Institute, University of Guelph, Guelph, Ontario N1G 2W1 Canada}
\date{\today}

\begin{abstract}
{The magneto-optical longitudinal, transverse Hall and circularly-polarized response of silicene and other materials described by a Kane-Mele Hamiltonian are calculated. Particular attention is paid to the effects of an external electric field and finite charge doping.  The energy of interband transitions can be tuned by varying the electric field.  The onset frequency of the absorptive peaks moves differently between the topological insulator and band insulator regimes.  This may be used to verify experimentally the existence of the two insulating phases as well as provide a measure of the spin-orbit band gap.  The zeroth Landau level splits between four spin and valley distinct energies resulting in valley-spin-polarized levels in the density of states.  With charge doping, transitions between these levels allow for a spin- and valley-polarized response in the conductivity whereby charge carriers of specific spin and valley index can be isolated by tuning the incident photon frequency.  Increasing the chemical potential is shown to redistribute spectral weight from interband transitions to a strong low-energy intraband response.  For large chemical potential, this intraband feature is associated with the semiclassical cyclotron resonance frequency which is shown to linearly increase with magnetic field.
}
\end{abstract}

\pacs{78.67.Wj, 78.20.Ls, 71.70.Di, 72.80.Vp
}

\maketitle

\section{Introduction}
\begin{figure}[h!]
\begin{center}
\includegraphics[width=1.0\linewidth]{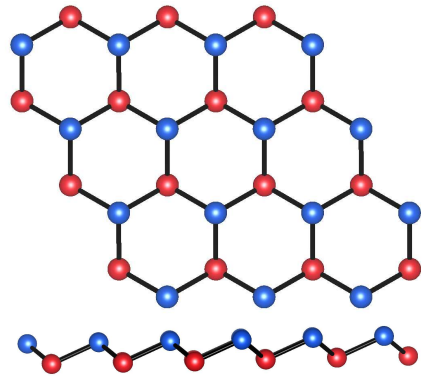}
\end{center}
\caption{\label{fig:Silicene}(Color online) Upper: The crystal structure of silicene is based on the 2D honeycomb lattice.  Due to the larger ionic size of silicon atoms, the silicene lattice is buckled with $A$ and $B$ sublattices sitting in vertically separated parallel planes.  Lower: Side view of the vertical buckling.
}
\end{figure}

 Two-dimensional (2D) crystals continue to attract experimental and theoretical attention as they remain a platform for investigating novel physics in addition to showing great promise for technological applications. While graphene, carbon atoms bonded together on a honeycomb lattice, was the first experimentally available 2D crystal, recent experiments\cite{Lalmi:2010, DePadova:2010, DePadova:2011, Vogt:2012, Lin:2012, Fleurence:2012} have been successful in synthesizing a monolayer of silicon atoms, known as silicene, which bond on a similar lattice.  This similarity results from carbon and silicon residing in the same column on the chemical periodic table. The larger ionic size of silicon atoms, however, causes an $sp^3$ hybridization in addition to the $sp^2$ hybridization found in carbon\cite{Drummond:2012}.  This mixture of hybridizations causes the 2D lattice of silicene to be buckled\cite{Liu:2011, Liu:2011a, Drummond:2012} such that sites on the $A$ and $B$ sublattices sit in different vertical planes with a separation of $d\approx 0.46$ \AA \cite{Ni:2012, Drummond:2012} as illustrated in Fig.~\ref{fig:Silicene}.  While, like graphene, the low-energy dynamics near the two valleys $K$ and $K^\prime$ of the hexagonal Brillouin zone are well described by Dirac theory, the electrons are massive due to a larger spin-orbit interaction\cite{Konschuh:2010}. This results in an energy gap in the band structure which is quoted as $\Delta_{\rm so}\approx 1.55-7.9$ meV from density functional theory\cite{Liu:2011, Liu:2011a, Drummond:2012} and tight-binding calculations\cite{Liu:2011a}.  As a result, it is argued that the Dirac nature of the fermions makes a topological insulator (TI).  It has been predicted that the application of an external electric field $E_z$ will create an on site potential difference ($\Delta_z=E_zd$) between sublattices yielding a tunable band gap\cite{Ni:2012, Drummond:2012, Ezawa:2012a}.  The electric field is also predicted to break the spin degeneracy at a given valley.  By varying the electric field such that $\Delta_z$ becomes greater than $\Delta_{\rm so}$, silicene is predicted to transition between a TI and a band insulator (BI)\cite{Drummond:2012,Ezawa:2012a}.  At the critical value $\Delta_z=\Delta_{\rm so}$, the lowest band gap at each valley closes forming a Dirac point.  This is referred to as a valley-spin-polarized metal (VSPM)\cite{Ezawa:2012}.  These exciting properties along with silicene's compatibility with the silicon dominated electronic industry make silicene a promising candidate for technological purposes.  Similar physics is predicted for germanene, the monolayer of germanium.  

An external magnetic field causes Landau levels to form in the electronic density of states. Transitions between these discrete levels generate absorption lines in the magneto-optical conductivity.  For graphene, this has been predicted theoretically\cite{Gusynin:2007, Gusynin:2007b, Pound:2012} and confirmed experimentally\cite{Sadowski:2006, Sadowski:2007, Jiang:2007, Deacon:2007, Plochocka:2008, Orlita:2010, Henriksen:2010, Orlita:2011}.  Recently, this has also been explored in other Dirac-like systems\cite{Tse:2011, Tabert:2013a, ZLi:2013, Ashby:2013}.  Unlike graphene, where the $n=0$ Landau level is pinned at $\varepsilon=0$\cite{Semenoff:1984, Gusynin:2007b, Gusynin:2007, Semenoff:2011, Pound:2012, Pound:2011, Pound:2011a}, the stronger spin-orbit coupling (SOC) in silicene causes the $n=0$ level to split between $\pm\Delta_{\rm so}/2$ much like the case of gapped graphene\cite{Gusynin:2007b, Gusynin:2007} which exists in the BI regime.  In the presence of an electric field, the single-valley Landau levels are no longer spin-degenerate and the onset of interband transitions can be controlled by tuning the electric field.

As there exist calculations on the zero magnetic field optical conductivity of silicene\cite{Stille:2012, Matthes:2013, Tabert:2013b} and only preliminary work on silicene in the presence of a magnetic field\cite{Tabert:2013a, Tahir:2013, Ezawa:2012c}, we examine the magneto-optical conductivity of silicene paying particular attention to the effects of the insulating phase (TI or BI) on both the real part of the longitudinal conductivity, the absorptive part of the transverse Hall conductivity and the response to circularly-polarized light.  This work builds on the results presented in Ref.~\cite{Tabert:2013a} where a special case of valley-spin-polarization is explored.  As Ref.~\cite{Tabert:2013a} is specific to a particular regime of charge doping and electric field, in this work, we examine the effect of varying the chemical potential $\mu$ and onsite potential difference.  We consider both charge neutrality and the system in the presence of a finite chemical potential; additionally, particular attention is given to the case of $\Delta_z=0$ which represents the results of a 2D Kane-Mele topological insulator.  A variety of electric field strengths are considered to elucidate signatures of the TI, BI and VSPM phases.  The semiclassical cyclotron resonance frequency is also investigated.  We also rehearse the results for the valley-spin-polarized DOS and circularly polarized light.  These results can be generalized to similar 2D crystals with sizable spin-orbit interactions such as germanene ($\Delta_{\rm so}\approx 24-93$ meV)\cite{Liu:2011, Liu:2011a}.

Our paper is organized as follows: in Sec.~II we give a theoretical background for silicene in a magnetic field.  In Sec.~III we examine the electronic density of states.  Sec.~IV contains results for the longitudinal magneto-optical conductivity.  The circularly-polarized response is presented in Sec.~V.  Sec.~VI features a discussion on the semiclassical cyclotron resonance frequency.  Our conclusions can be found in Sec.~VII.

 
\section{Theory for Silicene in an External Magnetic Field}

It has recently been shown that the low-energy physics of silicene is well approximated by a simple nearest-neighbour tight-binding Hamiltonian\cite{Liu:2011a, Ezawa:2012, Ezawa:2012a, Ezawa:2012b}.  Written about a single $K$ point, the effective low-energy Hamiltonian is\cite{Drummond:2012}
\begin{equation}
\hat{H}=v(\xi p_x\hat{\tau}_x+p_y\hat{\tau}_y)-\xi\frac{1}{2}\Delta_{\rm so}\sigma_z\tau_z+\frac{1}{2}\Delta_z\tau_z,
\end{equation}
where $\tau_i$ and $\sigma_i$ are the Pauli matrices associated with the pseudospin and real spin of the system, respectively. $\xi$ is the valley index for the two inequivalent $K$ points and can take the values of $\pm 1$ for the $K$ and $K^\prime$ points, respectively. $v\approx 5\times 10^5$m/s is the Fermi velocity and $p_x$ and $p_y$ are components of the momentum measured relative to the $K$ points.  The first term is the usual low-energy graphene-like Hamiltonian\cite{Neto:2009, Abergel:2010} for describing massless Dirac fermions.  The second term is of the Kane-Mele type\cite{Kane:2005} for intrinsic spin-orbit coupling with a spin-orbit band gap of $\Delta_{\rm so}$.  The final term is associated with the aforementioned sublattice potential difference due to the application of an external electric field\cite{Ezawa:2012, Ezawa:2012a, Ezawa:2012b, Drummond:2012}.    In Ref\cite{Ezawa:2012}, a Rashba SOC is also included; however, it is typically neglected\cite{Ezawa:2012} as it is an order of 10 smaller in magnitude than $\Delta_{\rm so}$.  Ignoring the Rashba term, the full 8x8 matrix spanning the two $K$ points is block diagonal in 2x2 matrices labelled by valley ($\xi=\pm 1$) and spin ($\sigma=\pm 1$ for up and down spin, respectively).  These 2x2 matrices are
\begin{equation}\label{Ham}
\hat{H}_{\xi\sigma}=\left(\begin{array}{cc}
-\frac{1}{2}\sigma\xi\Delta_{\rm so}+\frac{1}{2}\Delta_z & v(\xi p_x-ip_y)\\
v(\xi p_x+ip_y) & \frac{1}{2}\sigma\xi\Delta_{\rm so}-\frac{1}{2}\Delta_z
\end{array}\right).
\end{equation}

\begin{figure}[h!]
\begin{center}
\includegraphics[width=1.0\linewidth]{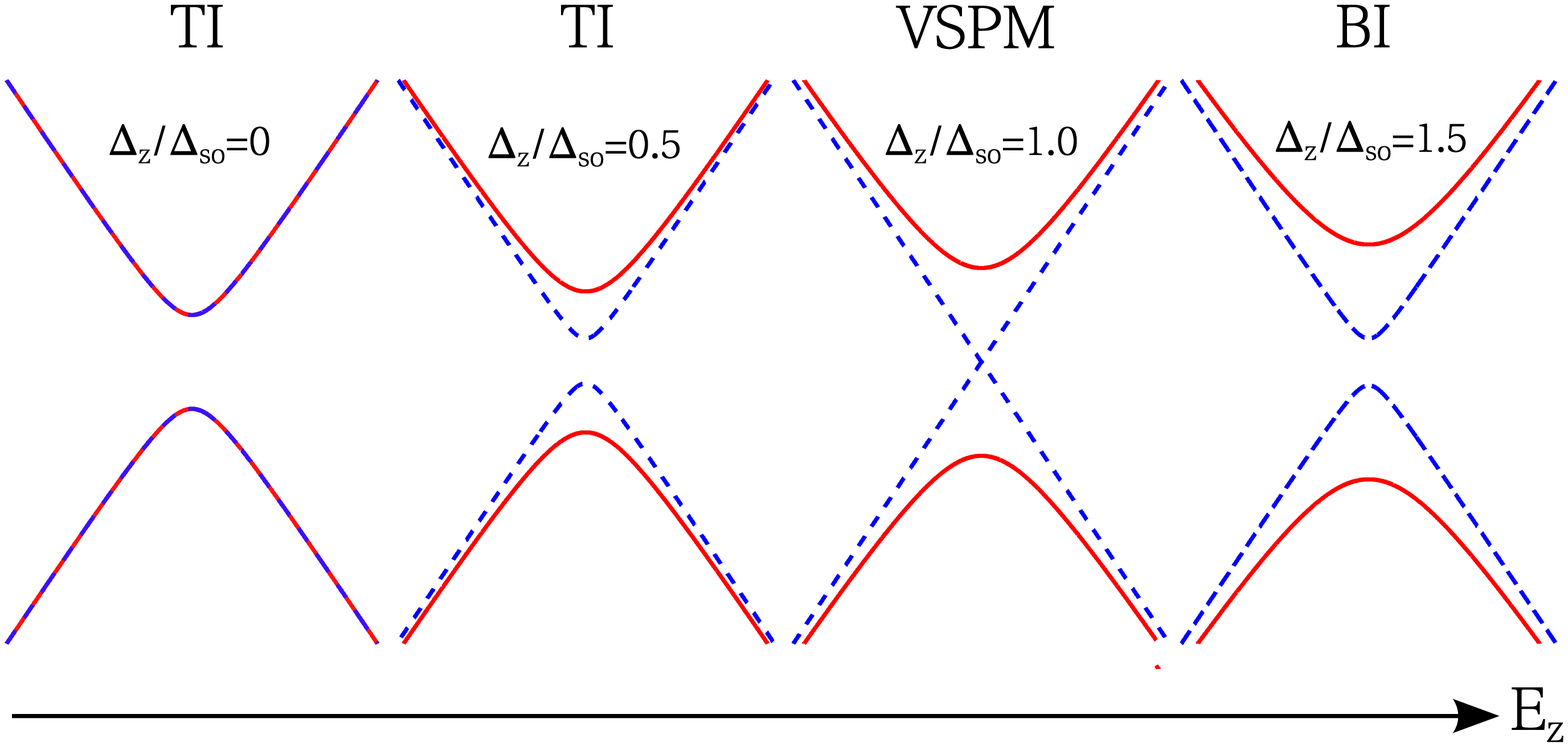}
\end{center}
\caption{\label{fig:Energy} (Color online) Schematic representation of the band structure evolution at the $K$ point as the perpendicular electric field strength is increased.  The dashed blue curves represent spin up bands and the solid red curves represent spin down bands.  A finite electric field spin splits the bands and the system transitions from a TI to a BI as $\Delta_z$ becomes greater than $\Delta_{\rm so}$.  When $\Delta_z=\Delta_{\rm so}$, the lowest band gap closes and the system is referred to as a VSPM.
}
\end{figure}

Solving Eqn.~\eqref{Ham}, gives the low-energy eigenvalues
\begin{equation}
\varepsilon_{\xi\sigma}=\pm\sqrt{\hbar^2v^2k^2+\Delta_{\xi\sigma}^2},
\end{equation}
where $\Delta_{\xi\sigma}=-\frac{1}{2}\sigma\xi\Delta_{\rm so}+\frac{1}{2}\Delta_z$.  A schematic plot of the band structure evolution about the $K$ point for increased $\Delta_z$ is shown in Fig.~\ref{fig:Energy}.  At the $K^\prime$ point, the spin labels switch.  When $\Delta_z=0$, all energy bands are spin-degenerate and separated by an insulating gap of $\Delta_{\rm so}$.  As the electric field is increased such that $\Delta_z<\Delta_{\rm so}$, the system remains a TI but the bands are spin split.  The system is now characterized by two energy gaps ($|\Delta_{K\uparrow}|$ and $|\Delta_{K\downarrow}|$).  In the TI regime, the lowest gap decreases with increased $\Delta_z$ while the second gap increases. When $\Delta_z=\Delta_{\rm so}$, the lowest band gap closes.  As $\Delta_z$ is increased further, the system transitions into the BI regime and the lowest energy gap reopens; although, a band inversion has occurred\cite{Ezawa:2012a} associated with a change in pseudospin label of the two lowest gapped bands.  In this regime, both gaps increase with increased $\Delta_z$.  
The evolution of the band structure plays an important role in the magneto-optics of silicene.

\begin{figure}[h!]
\begin{center}
\includegraphics[width=1.0\linewidth]{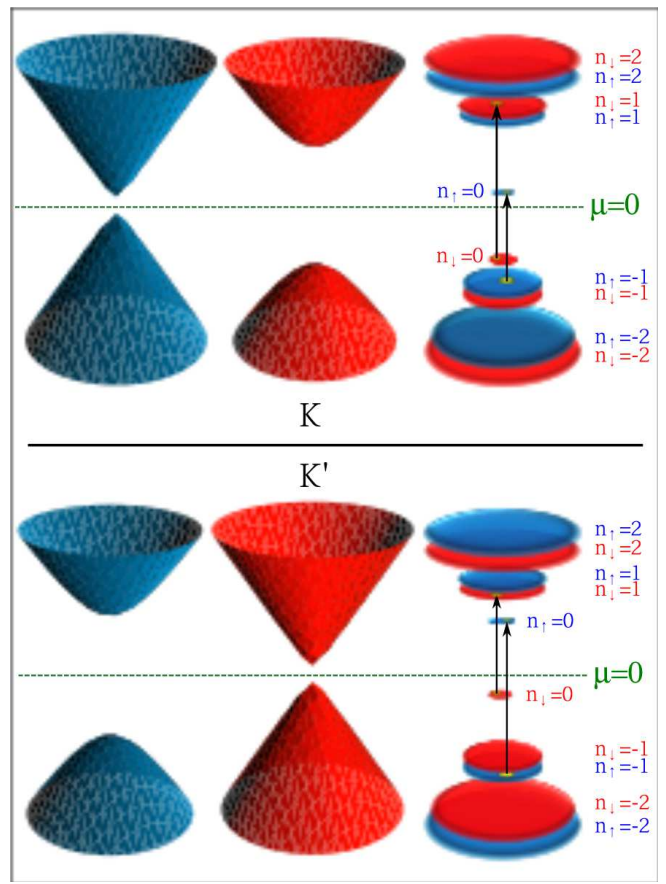}
\end{center}
\caption{\label{fig:LL-3D} (Color online) Schematic representation of the Landau level formation in the TI regime about the $K$ and $K^\prime$ points (upper and lower frames, respectively).  The blue bands shown to the left are spin up while the red bands seen in the middle are spin down.  These bands are associated with no magnetic field and are separated for clarity.  The Landau levels are shown to the right and are not spin-degenerate at a given valley.  All but the $n=0$ levels have an equal-energy counterpart of opposite spin label at the other valley.  The Fermi energy at charge neutrality (\emph{i.e.} chemical potential $\mu=0$) is given by the dashed green line.  The two lowest energy interband transitions are given by the black arrows.
}
\end{figure}

Now, consider the effect of an external magnetic field of strength $B$ which, by choice, is oriented in the $z$ direction.  Working in the Landau gauge, the vector potential $\bm{A}=\nabla \times\bm{B}$, is written as $\bm{A}=(-By,0,0)$.  The magnetic field changes the momentum operators by the usual Peierls substitution
\begin{equation}
\hat{p}_{i}\rightarrow\hat{p}_{i}+\frac{q}{c}\hat{A}_i,
\end{equation}
where $q=e$ is the elementary charge and $c$ is the speed of light.  Therefore, the low-energy Hamiltonian becomes
\begin{align}\label{HamB}
\hat{H}_{\sigma\xi}=\left(\begin{array}{cc}
-\frac{1}{2}\sigma\xi\Delta_{\rm so}+\frac{1}{2}\Delta_z & v\left[\xi\left(p_x-\frac{e}{c}By\right)-ip_y\right]\\
v\left[\xi\left(p_x-\frac{e}{c}By\right)+ip_y\right] & \frac{1}{2}\sigma\xi\Delta_{\rm so}-\frac{1}{2}\Delta_z
\end{array}\right).
\end{align}
Solving the simple eigenvalue equation $\hat{H}\psi=\varepsilon\psi$, where $\psi^T=(\phi_A,\phi_B)$ is the wavefunction with components associated with the $A$ and $B$ sublattices, gives the Landau level spectrum
\begin{equation}
\varepsilon_n=\text{sgn}(n)\sqrt{\Delta_{\xi\sigma}^2+2|n|v^2\frac{\hbar e B}{c}},
\end{equation}\label{Energy}
where $n=0,\pm 1,\pm 2,...$ is the Landau level index.  A schematic representation of the Landau level formation for finite $\Delta_z<\Delta_{\rm so}$ at the $K$ and $K^\prime$ points is shown in the upper and lower frames of Fig.\ref{fig:LL-3D}, respectively. 
The blue bands represent spin up while the red bands are spin down.  The same color scheme applies for the Landau levels shown to the right.  The spin bands shown on the left and middle for $B=0$ are separated for clarity.  For finite $\Delta_z$, the Landau levels at a given valley are spin-split; however, in the total system, all but the $n=0$ levels are spin-degenerate.

In principle, there is also a Zeeman interaction which shifts the Landau levels by energy $\varepsilon_z\approx 2\mu_BB$ with $\mu_B\approx 5.78\times 10^{-2}$meV/T the Bohr magneton.  For a magnetic field of 1T, this effect shifts the spin levels by $\varepsilon_z\approx 0.1$meV and it is therefore ignored\cite{Tahir:2012, Ezawa:2012, Ezawa:2012a, Ezawa:2012b, Tabert:2013a} as it will have a negligible effect on the conductivity.  

\begin{figure}[h!]
\begin{center}
\includegraphics[width=1.0\linewidth]{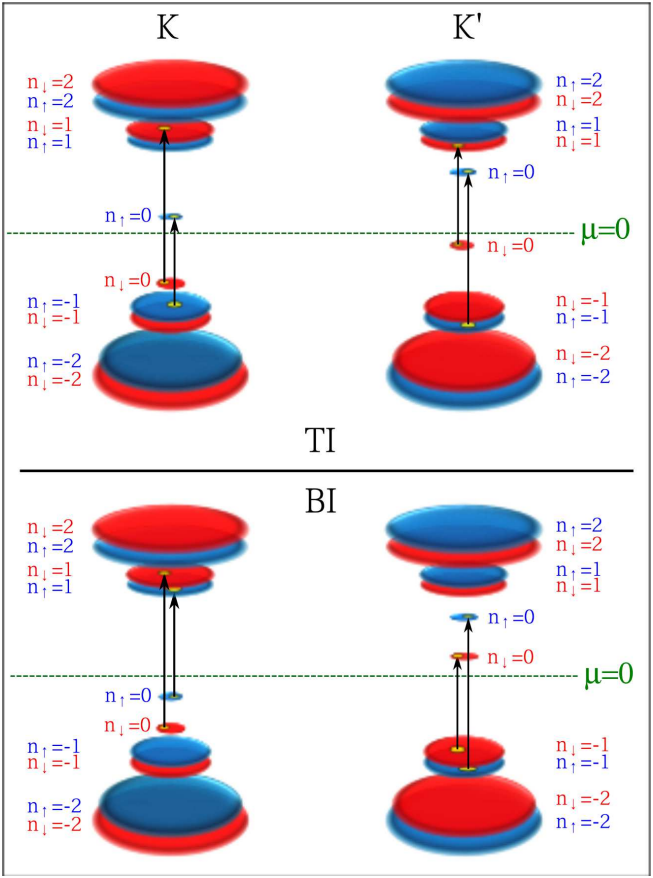}
\end{center}
\caption{\label{fig:LL} (Color online) Schematic representation of the Landau levels in the TI (upper frame) and BI (lower frame) regimes about the $K$ and $K^\prime$ points.  Again, blue and red signify spin up and down, respectively.  While the relative position of the $n\neq 0$ levels remain the same, the $n=0$ level undergoes an important transition due to the band inversion.  That is, the $n=0$ spin up level at the $K$ point switches between positive and negative energy while the $n=0$ spin down level switches between negative and positive energy at the $K^\prime$ point.
}
\end{figure}

The $n=0$ Landau level must be treated carefully and is given by $\varepsilon_0=-\xi\Delta_{\xi\sigma}$.  The zeroth Landau level undergoes an important transition between the TI ($\Delta_z<\Delta_{\rm so}$) and BI ($\Delta_z>\Delta_{\rm so}$) regimes which are shown in the upper and lower frames of Fig.~\ref{fig:LL}, respectively. 
Given the expressions for $\Delta_{\xi\sigma}$ and $\varepsilon_0$, it is evident that in the TI regime, the $n=0$ spin up Landau levels are at positive energy while the $n=0$ spin down levels are at negative energy, refer to the upper frame of Fig.~\ref{fig:LL}. $\mu=0$ represents the location of zero energy.  In the BI regime, $\Delta_{z}$ is greater than $\Delta_{\rm so}$ resulting in a change in sign of $\Delta_{K\uparrow}$ and $\Delta_{K^\prime\downarrow}$ (and, therefore, $\varepsilon_0(\Delta_{K\uparrow})$ and $\varepsilon_0(\Delta_{K^\prime\downarrow})$).  Thus, the $n=0$ spin up level at the $K$ point and the $n=0$ spin down level at $K^\prime$ switch sign.  This is a signature of the aforementioned band inversion associated with the transition between the TI and BI regimes.  In the VSPM state (not shown) the $n=0$ spin up (down) level at the $K$ ($K^\prime$) point sits at zero energy.   

\begin{figure}[h!]
\begin{center}
\includegraphics[width=1.0\linewidth]{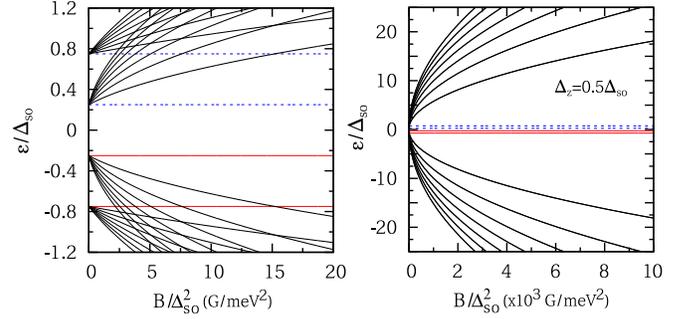}
\end{center}
\caption{\label{fig:Fan} (Color online) Landau level energies as a function of magnetic field for $\Delta_{z}=0.5\Delta_{\rm so}$.  All $n\neq 0$ levels scale as $\sqrt{B}$ while the location of the $n=0$ levels are fixed by the strength of $\Delta_{\rm so}$ and $\Delta_z$.  The left frame shows low magnetic field results while the right frame illustrates the evolution over a large magnetic field range (note the factor of $10^3$).
}
\end{figure}

A plot of the Landau level evolution with varying magnetic field for $\Delta_z=0.5\Delta_{\rm so}$ is shown in Fig.~\ref{fig:Fan}.   
Similar to graphene, the $n\neq 0$ levels scale as $\sqrt{B}$.  Unlike graphene, the $n=0$ level is not pinned at zero energy.  The four spin- and valley-split $n=0$ levels do not scale with the magnetic field but are only controlled through the spin-orbit interaction and perpendicular electric field.  The separation between $n=0$ levels at different valleys is adjusted by tuning the electric field; however, the separation between $n=0$ levels at the same valley is fixed at $\Delta_{\rm so}$.  For low magnetic field (left frame), two of the $n=0$ levels are higher in energy than several higher numbered levels.  This behaviour becomes important in the magneto-optical conductivity when trying to tune the onset of interband transitions.

The wavefunction solutions to the eigenvalue equation are required when calculating the magneto-optical conductivity tensor.  Using Eqn.\eqref{HamB}, the eigenvectors are found to be
\begin{equation}
\left|\bar{n}\right\rangle_K=
\left(\begin{array}{c}
-iA_{n}\left||n|-1\right\rangle\\
B_{n}\left||n|\right\rangle
\end{array}\right)
\end{equation}
and
\begin{equation}
\left|\bar{n}\right\rangle_{K^\prime}=
\left(\begin{array}{c}
-iA_{n}\left||n|\right\rangle\\
B_{n}\left||n|-1\right\rangle
\end{array}\right),
\end{equation}
where $\left||n|\right\rangle$ is an orthonormal Fock state of the harmonic oscillator and
\begin{equation}\label{An}
A_{n}=\left\lbrace\begin{array}{cc}
\displaystyle\frac{\text{sgn}(n)\sqrt{|\varepsilon_n|+\text{sgn}(n)\Delta_{\xi\sigma}}}{\sqrt{2|\varepsilon_n|}}, &\quad n\neq 0,\\
\displaystyle\frac{1-\xi}{2}, &\quad n=0,
\end{array}\right.
\end{equation}
and
\begin{equation}\label{Bn}
B_{n}=\left\lbrace\begin{array}{cc}
\displaystyle\frac{\sqrt{|\varepsilon_n|-\text{sgn}(n)\Delta_{\xi\sigma}}}{\sqrt{2|\varepsilon_n|}}, &\quad n\neq 0,\\
\displaystyle\frac{1+\xi}{2}, &\quad n=0.
\end{array}\right.
\end{equation}
These wavefunctions and corresponding eigenenergies reduce to those of gapped graphene\cite{Gusynin:2006, Gusynin:2007, Gusynin:2007b} with the substitution $\Delta_{\xi\sigma}\rightarrow\Delta$, placing the system in the BI regime.  

With the wavefunctions and energy dispersion, the magneto-optical conductivity is found through the Kubo formula\cite{Mahan:1990} in the usual way\cite{Tse:2011, Tabert:2013a}
\begin{equation}\label{Cond}
\sigma_{\alpha\beta}(\Omega)=\frac{ieB}{2\pi c}\sum_{\sigma=\pm 1}\sum_{\xi=\pm 1}\sum_{n m} \frac{f_m-f_n}{\varepsilon_n-\varepsilon_m}\frac{\left\langle \bar{m}\right|\hat{j}_\alpha\left|\bar{n}\right\rangle\left\langle \bar{n}\right|\hat{j}_\beta\left|\bar{m}\right\rangle}{\hbar\Omega+\varepsilon_m-\varepsilon_n+i\eta},
\end{equation}
where $f_m=1/[1+\text{exp}(\beta(\varepsilon_m-\mu))]$ is the Fermi distribution function with $\beta=1/(k_BT)$, $\varepsilon_m$ is the energy of the $m^{th}$ Landau level, $\hat{j}_{\alpha}=ev\hat{\tau}_{\alpha}$ is the current operator with $\tau_{\alpha}$ ($\alpha=x,y$) the usual Pauli matrices and $n,m$ index over all Landau level sites.  Here, $\eta$ will represent a phenomenological transport scattering rate taken to be constant. The familiar selection rules\cite{Gusynin:2007, Gusynin:2007b, Tse:2011} $|n|=|m|\pm 1$ for Landau level transitions are found through an evaluation of the matrix elements.  In the zero temperature limit, the Fermi functions can be replaced by step functions.  In what follows, we assume a positive value of $\mu$ and drop the absolute value signs on $n$ as we assume all transitions to negative Landau levels are Pauli blocked.  This can easily be adjusted by including $|n|$ in the Kronecker $\delta$-functions of the conductivity formulas.  Therefore, the real and imaginary parts of the zero temperature longitudinal ($xx$) magneto-optical conductivity are\cite{Tabert:2013a} 
\begin{align}\label{Condxx}
\frac{\text{Re}\sigma_{xx}(\Omega)}{\sigma_0}&=\frac{2v^2\hbar e B}{\pi c}\sum_{\sigma=\pm 1}\sum_{\xi=\pm 1}\sum_{n,m}\frac{\Theta(\mu-\varepsilon_m)-\Theta(\mu-\varepsilon_n)}{\varepsilon_n-\varepsilon_m}\notag\\
&\times\left[\left(A_{m}B_{n}\right)^2\delta_{n,|m|-\xi}
+\left(B_{m}A_{n}\right)^2\delta_{n,|m|+\xi}\right]\notag\\
&\times\frac{\eta}{\eta^2+(\hbar\Omega+\varepsilon_m-\varepsilon_n)^2},
\end{align}
and
\begin{align}\label{Condxx-i}
\frac{\text{Im}\sigma_{xx}(\Omega)}{\sigma_0}&=\frac{2v^2\hbar e B}{\pi c}\sum_{\sigma=\pm 1}\sum_{\xi=\pm 1}\sum_{n,m}\frac{\Theta(\mu-\varepsilon_m)-\Theta(\mu-\varepsilon_n)}{\varepsilon_n-\varepsilon_m}\notag\\
&\times\left[\left(A_{m}B_{n}\right)^2\delta_{n,|m|-\xi}
+\left(B_{m}A_{n}\right)^2\delta_{n,|m|+\xi}\right]\notag\\
&\times\frac{\hbar\Omega+\varepsilon_m-\varepsilon_n}{\eta^2+(\hbar\Omega+\varepsilon_m-\varepsilon_n)^2},
\end{align}
respectively, where $\sigma_0=e^2/(4\hbar)$ and $\Theta(x)$ is the Heaviside step function which enforces the Pauli exclusion principle for optical transitions, \emph{i.e.}, transitions can only occur between an occupied ($m$) and an unoccupied ($n$) state.  The corresponding equations for the real and imaginary parts of the transverse Hall conductivity are\cite{Tabert:2013a}
\begin{align}\label{Condxy-Re}
\frac{\text{Re}\sigma_{xy}(\Omega)}{\sigma_0}&=-\frac{2v^2\hbar e B}{\pi c}\sum_{\sigma=\pm 1}\sum_{\xi=\pm 1}\sum_{n,m}\xi\frac{\Theta(\mu-\varepsilon_m)-\Theta(\mu-\varepsilon_n)}{\varepsilon_n-\varepsilon_m}\notag\\
&\times\left[\left(A_{m}B_{n}\right)^2\delta_{n,|m|-\xi}
-\left(B_{m}A_{n}\right)^2\delta_{n,|m|+\xi}\right]\notag\\
&\times\frac{\hbar\Omega+\varepsilon_m-\varepsilon_n}{\eta^2+(\hbar\Omega+\varepsilon_m-\varepsilon_n)^2},
\end{align}
and
\begin{align}\label{Condxy}
\frac{\text{Im}\sigma_{xy}(\Omega)}{\sigma_0}&=\frac{2v^2\hbar e B}{\pi c}\sum_{\sigma=\pm 1}\sum_{\xi=\pm 1}\sum_{n,m}\xi\frac{\Theta(\mu-\varepsilon_m)-\Theta(\mu-\varepsilon_n)}{\varepsilon_n-\varepsilon_m}\notag\\
&\times\left[\left(A_{m}B_{n}\right)^2\delta_{n,|m|-\xi}
-\left(B_{m}A_{n}\right)^2\delta_{n,|m|+\xi}\right]\notag\\
&\times\frac{\eta}{\eta^2+(\hbar\Omega+\varepsilon_m-\varepsilon_n)^2},
\end{align}
respectively.  Note that Re$\sigma_{xx}(\Omega)$ and Im$\sigma_{xy}(\Omega)$ correspond to the absorptive parts of the longitudinal and transverse Hall conductivities, respectively, which means that the absorption peaks shown here will appear as dips in the experimentally measured transmission\cite{Orlita:2010}.

\section{Density of States}

\begin{figure}[h!]
\begin{center}
\includegraphics[width=1.0\linewidth]{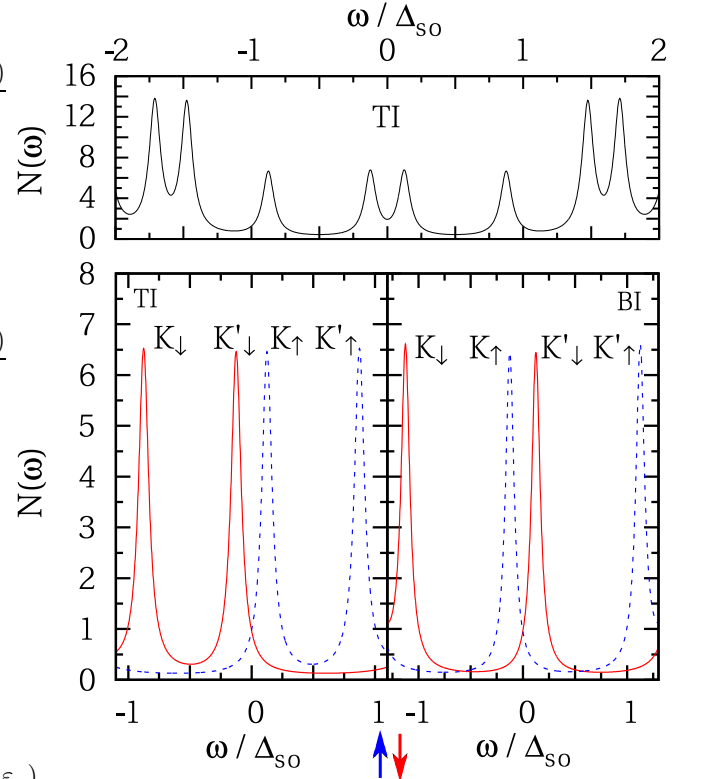}
\end{center}
\caption{\label{fig:DOS}(Color online) Upper: Total electronic density of states for silicene in a magnetic field in units of\cite{Note} $eB/(2\pi\hbar c\Delta_{\rm so})$.  Lower: Spin dependent density of states in the TI (left) and BI (right) regimes for $\Delta_z=0.75\Delta_{\rm so}$ and $1.25\Delta_{\rm so}$, respectively.  Dashed blue curves correspond to spin up and the solid red curves correspond to spin down.  The levels associated with $n=0$ are spin- and valley-polarized while all higher features are spin-valley-degenerate and of double intensity.
}
\end{figure}

The low-energy density of states for silicene in a magnetic field is calculated by the relation
\begin{equation}
N(\omega)=\frac{eB}{2\pi\hbar c}\sum_{\sigma=\pm 1}\sum_{\xi=\pm 1}\sum_{n=-\infty}^{\infty}\delta(\omega-\varepsilon_n).
\end{equation}
Using the Lorentzian representation of the $\delta$-function, $\delta(x)\rightarrow(\eta/\pi)/(\eta^2+x^2)$, the density of states can be evaluated\cite{Tabert:2013a}.  A plot of the total electronic density of states is shown in the upper frame of Fig.~\ref{fig:DOS}.
The spin dependent density of states is shown in the lower two frames for the TI (left) and BI (right) regimes for $\Delta_z=0.75\Delta_{\rm so}$ and $1.25\Delta_{\rm so}$, respectively.  In all cases, $B/\Delta_{\rm so}^2=65.7$G/meV$^2$ and a scattering rate of $\eta=0.05\Delta_{\rm so}$ has been used.  Four spin- and valley-polarized levels are located at $\omega=-\Delta_{K\downarrow}$, $\Delta_{K^\prime\downarrow}$, $-\Delta_{K\uparrow}$ and $\Delta_{K^\prime\uparrow}$ corresponding to the four $n=0$ Landau levels.  The two lower frames illustrate the shift in two of the $n=0$ Landau levels that results from the band inversion associated with the transition from the TI to BI regime.  While the $n=0$ levels are spin- and valley-polarized, all higher features are spin- and valley-degenerate.  Thus, all $n\neq 0$ levels are of double weight.  As Landau levels in graphene have been detected experimentally\cite{Li:2007, Miller:2009, Andrei:2012}, the four low-energy spin- and valley-polarized levels in silicene should be observed by scanning tunneling spectroscopy.

\section{Results for the Magneto-Optical Conductivity}

\begin{figure}[h!]
\begin{center}
\includegraphics[width=1.0\linewidth]{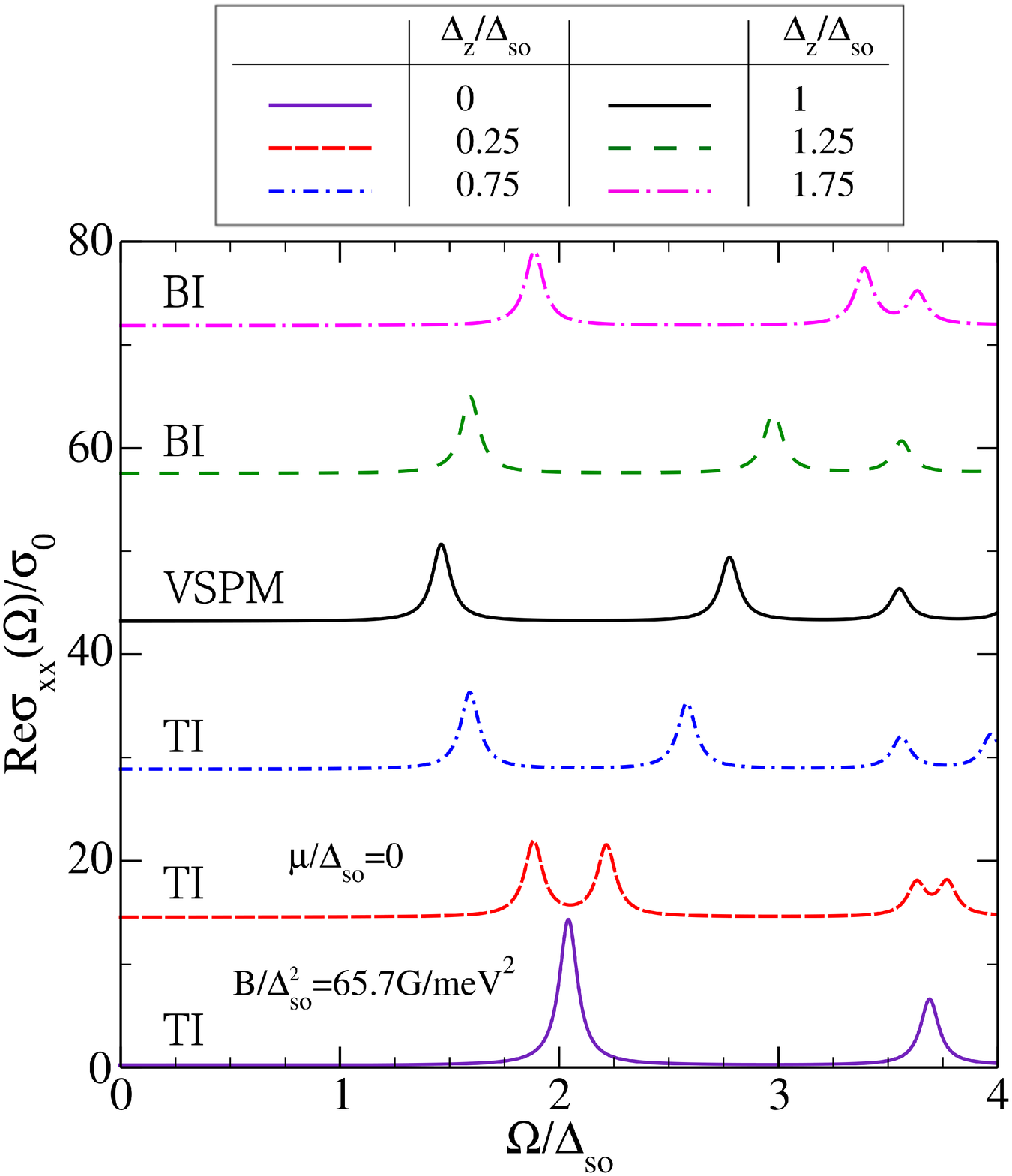}
\end{center}
\caption{\label{fig:Cond-Delta-mu0}(Color online) Real part of the zero temperature longitudinal magneto-optical conductivity of silicene for varying electric field strength in a magnetic field of strength $B/\Delta_{\rm so}^2=65.7$G/meV$^2$ with a scattering rate of $\eta=0.05\Delta_{\rm so}$ and $\mu=0$.  The behaviour of the interband responses is a signature of the two insulating regimes.   The results are vertically offset by 15 units.
}
\end{figure}

An evaluation of Eqn.~\eqref{Condxx} for the absorptive part of the longitudinal magneto-optical conductivity is shown in Fig.~\ref{fig:Cond-Delta-mu0} for varying $\Delta_z$ at charge neutrality ($\mu=0$).  In reference to the forthcoming discussion, an interband transition refers to a transition between Landau levels which arise from different $B=0$ bands.  Likewise, an intraband transition is one between levels from the same $B=0$ band.
For $\Delta_z=0$ (lowest purple curve) there are strong absorptive responses associated with interband transitions subject to the selection rules.  The energy of the first feature is set by the difference in energy of the $n=0$ and 1 Landau levels, that is, $\Omega=\Delta_{\rm so}+\varepsilon_{1}(\Delta_z=0)$.  As $\Delta_z$ is increased, each interband feature splits in two as a result of all Landau levels becoming spin split.  The intensity of the peaks is reduced due to a redistribution of spectral weight between the features.  The lowest of the split peaks, moves to lower energy as $\Delta_z$ is increased which is a signature of the closing of the lowest band gap of the $B=0$ bands.  The second split peak moves higher in energy due to the second band gap increasing.  When $\Delta_z=\Delta_{\rm so}$, the first feature is set by $\Omega=\varepsilon_1(\Delta_{K\uparrow})$, or equivalently, $\Omega=\varepsilon_1(\Delta_{K^\prime\downarrow})$.  As the system transitions through the VSPM state into the BI regime, all interband features move to higher energy due to the lowest gap reopening.  The variation in interband onset is a signature of the two insulating regimes and should allow for an experimental verification of their existence.

\begin{figure}[h!]
\begin{center}
\includegraphics[width=1.0\linewidth]{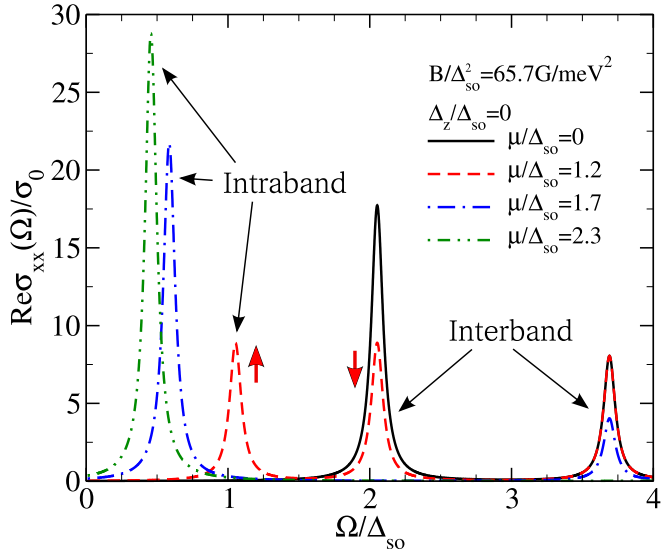}
\end{center}
\caption{\label{fig:Cond-Mu-D0}(Color online) Real part of the zero temperature longitudinal conductivity of silicene ($\Delta_z=0$) for varying chemical potential in a magnetic field of strength $B/\Delta_{\rm so}^2=65.7$G/meV$^2$ with a scattering rate of $\eta=0.05\Delta_{\rm so}$.   With increased $\mu$, the spectral weight of interband transitions is redistributed to a single intraband feature.  The two red arrows mark the spin-polarization of the two lowest features of the dashed red curve.  These results correspond to the Kane-Mele model for a 2D quantum-spin-hall insulator.
}
\end{figure}

\begin{figure}[h!]
\begin{center}
\includegraphics[width=0.830\linewidth]{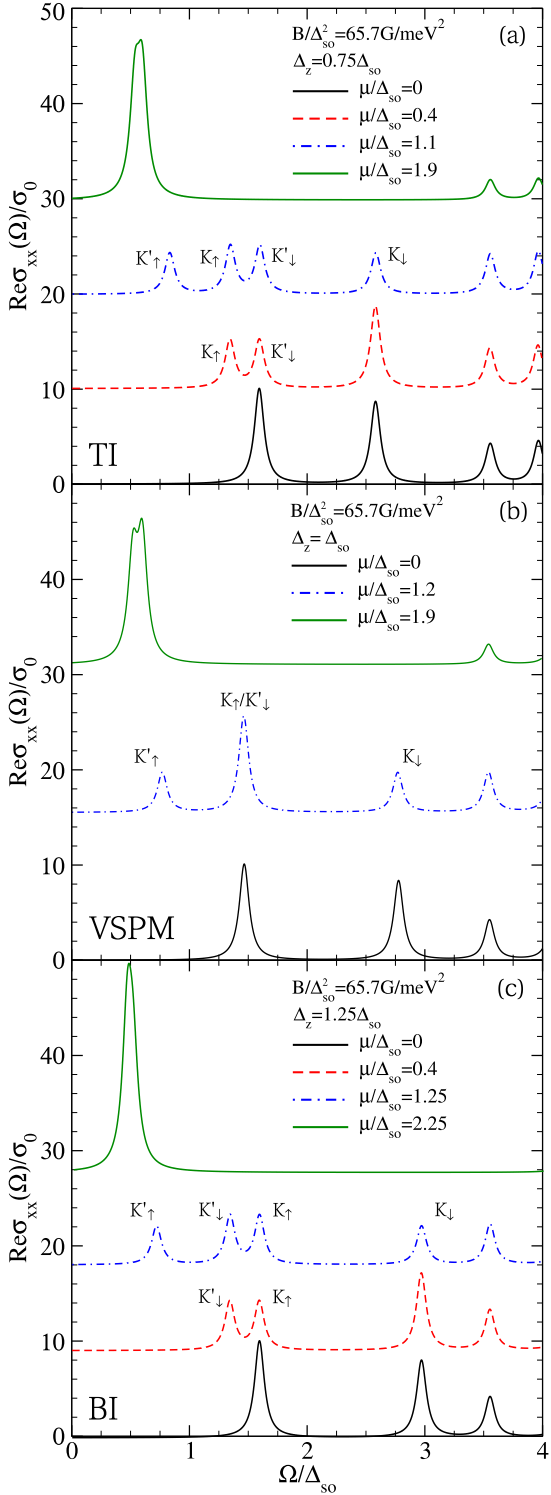}
\end{center}
\caption{\label{fig:Cond-mu}(Color online) Real part of the zero temperature longitudinal conductivity of silicene for varying chemical potential for $B/\Delta_{\rm so}^2=65.7$G/meV$^2$ and $\eta=0.05\Delta_{\rm so}$ in the (a) TI, (b) VSPM and (c) BI regimes.  Four spin- and valley-polarized responses can be generated. The results are vertically offset by (a) 10 units, (b) 16 and (c) 9 units.
}
\end{figure}

The effect of varying the chemical potential when $\Delta_z=0$ is shown in Fig.~\ref{fig:Cond-Mu-D0}.  This case represents the results of the Kane-Mele Hamiltonian for a quantum spin Hall insulator and 2D topological insulator.  As $\Delta_z=0$ corresponds to a topological insulator (two spin-split $n=0$ Landau levels at each valley), these results differ from those shown in Ref.~\cite{Gusynin:2007, Gusynin:2007b} for graphene with an asymmetry gap as that system is a band insulator (one spin-degenerate $n=0$ Landau level per valley).  This difference affects the spin and valley contributions to the low-energy features; however, the qualitative results of Ref.~\cite{Gusynin:2007, Gusynin:2007b} are retained.  Including a finite chemical potential such that the Fermi energy lies between all the $n=0$ and 1 Landau levels (dashed red curve), causes the lowest interband feature of the $\mu=0$ system (solid black curve) to decrease in intensity by a factor of two as half the spectral weight is shifted to a low-energy intraband peak\cite{Gusynin:2007, Gusynin:2007b}.  However, as opposed to Ref.~\cite{Gusynin:2007b}, where the two lowest peaks are valley-polarized, here the two lowest (red) features are spin-polarized as marked by the red arrows.  When $\mu$ is situated between the $n=1$ and 2 levels (dash-dotted blue curve), the interband absorption peak associated with transitions to and from the $n=0$ levels disappears due to Pauli blocking.  The spectral weight of the next highest response is diminished as interband transitions to the $n=1$ level are Pauli blocked.   With higher $\mu$, a similar redistribution of interband to intraband spectral weight is observed.  The low energy intraband signature moves to lower energy as a result of the $\sqrt{n}$ spacing between Landau levels which causes adjacent levels to be closer together at higher energy.

\begin{figure}[h!]
\begin{center}
\includegraphics[width=1.0\linewidth]{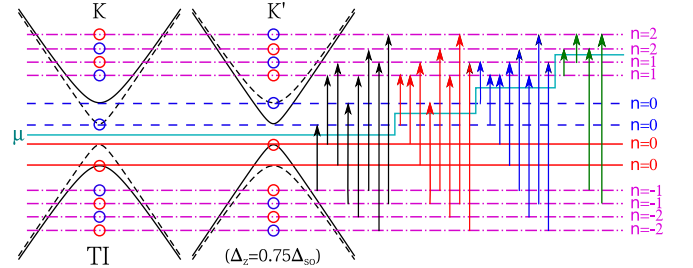}
\end{center}
\caption{\label{fig:Trans}(Color online) Schematic of the allowed transitions between Landau levels at both valleys in the TI regime for various values of chemical potential.  Dashed blue lines and circles represent spin up levels, solid red lines and circles represent spin down levels and dash dotted purple lines represent spin-degenerate levels.  The various chemical potential values used in Fig.~\ref{fig:Cond-mu}(a) are marked by the solid cyan line.
}
\end{figure}

The effect of varying the chemical potential for finite $\Delta_z$ is shown in Fig.~\ref{fig:Cond-mu}(a), (b) and (c) for the TI ($\Delta_z/\Delta_{\rm so}=0.75$), VSPM ($\Delta_z/\Delta_{\rm so}=1$) and BI ($\Delta_z/\Delta_{\rm so}=1.25$) regimes, respectively.  A finite $\Delta_z$ spin splits the Landau levels which allows for a much richer structure in the magneto-optical conductivity.  Similar to the case of $\Delta_z=0$, moving the Fermi energy causes a redistribution of spectral weight from interband to intraband responses.  The low-energy transitions which yield the results of Fig.~\ref{fig:Cond-mu}(a) are marked by arrows in Fig.~\ref{fig:Trans}.
The Landau levels at each valley are represented by circles colored blue and red for spin up and down, respectively.  Horizontal lines are drawn between the two valleys to signify the total Landau level contribution in the system.  These lines are dashed blue, solid red and dashed dotted purple lines for spin up, spin down and spin-degenerate levels, respectively.  The Landau level index is marked on the far right and the various values of chemical potential are given by the solid cyan line.  The lowest energy transitions for each value of chemical potential are represented by colored arrows where transitions are only permitted between like spin states.  The arrows are ordered by increasing length such that a change in length corresponds to a new absorption feature in the conductivity.  Thus, in reference to Fig.~\ref{fig:Cond-mu}(a), the first peak of the solid black curve is associated with the first two black arrows in Fig.~\ref{fig:Trans}, and the second, third and fourth features are likewise represented by the second, third and fourth pairs of black arrows, respectively.  For the dashed red curve, the first red arrow yields the first feature while the second arrow gives rise to the second peak.  The next absorption line results from the next two arrows while the remaining pairs of arrows yield the final two features, respectively.  With respect to the dash-dotted blue curve, the first four blue arrows result in the first four peaks, respectively, while the next two features come from the reaming two pairs of arrows.  The strong low energy signature of the solid green curve comes from the first two arrows with the two higher energy peaks being represented by the remaining pairs of arrows, respectively.

Returning to Fig.~\ref{fig:Cond-mu}(a), when $\mu$ is placed between the $n=0$ spin up levels at the $K$ and $K^\prime$ points (dashed red curve) such that it is above three $n=0$ levels, \emph{i.e.} $|\Delta_{K\uparrow}|<\mu<|\Delta_{K^\prime\uparrow}|$, the lowest interband feature redistributes its spectral weight between itself and a low energy intraband peak.  These intraband transitions result entirely from spin up electrons at the $K$ point while the remaining interband response is associated entirely with spin down electrons at $K^\prime$.  If $\mu$ is now situated between all the $n=0$ and 1 Landau levels, \emph{i.e.} $|\Delta_{K^\prime\uparrow}|<\mu<\varepsilon_1(\Delta_{K\uparrow})$, the second interband feature redistributes its spectral weight into a lower-energy intraband signature.  These intraband transitions result entirely from spin up electrons at $K^\prime$ and the remaining interband transitions are associated with spin down at $K$.  The other two spin- and valley-polarized responses remain and, for this doping, there are four robust spin- and valley-polarized peaks making it is possible to generate charge carriers of definite spin and valley label\cite{Tabert:2013a}.  The two lowest features are associated with intraband transitions while the upper two result from interband transitions.  The spin- and valley-polarized responses onset at $\Omega=\varepsilon_1(\Delta_{K^\prime\uparrow})-\Delta_{K^\prime\uparrow}$, $\varepsilon_1(\Delta_{K^\prime\downarrow})-\Delta_{K^\prime\downarrow}$, $\varepsilon_1(\Delta_{K\uparrow})-\Delta_{K\uparrow}$ and $\varepsilon_1(\Delta_{K\downarrow})-\Delta_{K\downarrow}$ for spin up and down at $K^\prime$ and spin up and down at $K$, respectively. While valley-spin polarization is predicted in the response to circularly-polarized light\cite{Ezawa:2012b, Stille:2012}, it is limited to only two pure spin-valley-polarized species which can only be selected by changing the insulating regime or handedness of the polarized light\cite{Stille:2012, Zhou:2012}.  Here, robust valley-spin polarization is present in both insulating regimes even in the longitudinal response\cite{Tabert:2013a} allowing any spin-valley-polarized response to be isolated by tuning the incident photon frequency. A more detailed discussion is found in Ref.~\cite{Tabert:2013a}. As $\mu$ is increased further, only one intraband signature remains which is associated with the semiclassical cyclotron resonance frequency (see Sec.~VI). The spectral weight of this feature increases with increasing $\mu$.

In the VSPM phase (Fig.~\ref{fig:Cond-mu}(b)) the potential for four spin- and valley-polarized responses is no longer present as two of the $n=0$ Landau levels are now at zero energy due to the closing of the lowest gap of the $B=0$ bands.  The conductivity curves look similar to the results for the TI regime; however, when the chemical potential is placed between all the $n=0$ and 1 Landau levels (dash-dotted blue curve), there are only two valley-spin-polarized responses.  For this value of chemical potential, the lowest feature of the $\mu=0$ system remains and is made of an equal mixture of spin and valley species.  This absorption signature is associated equally with interband and intraband transitions.  By examining the Landau levels which contribute to the degenerate $n=0$ level (see the upper frame of Fig.~\ref{fig:LL}) it is apparent that the spin up contribution is associated with an intraband transition while the spin down piece results from an interband transition.  This is similar to the behaviour of the $n=0$ Landau level in graphene\cite{Gusynin:2007}.  Again, there is a spectral weight redistribution to a strong intraband response for increased $\mu$.  A double peak feature is present in the strong intraband response as seen by the lowest feature in the solid green curves of Fig.~\ref{fig:Cond-mu}.  This results from the Landau levels being spin split and thus intraband transitions between different spin levels at a given valley are not of equal energy.  The separation between the double peaks decreases as $\mu$ increases.

The conductivity in the BI regime (see Fig.~\ref{fig:Cond-mu}(c)) is analogous to that of the TI regime; however, in this case, the spin and valley labels of the two middle spin- and valley-polarized responses switch due to the band inversion.  While the relative onset inverts, the separation between the two middle peaks always remains at $2|\Delta_{K\uparrow}|$ for $\Delta_z\neq 0$ and, thus, decreases with increasing $E_z$ in the TI regime until the polarization is lost in the VSPM state.  The separation then increases with increasing $E_z$ in the BI regime, however, the spin and valley labels switch.  Aside from the four spin- and valley-polarized peaks, all other transitions are made of an equal spin and valley mixture.

The onset frequency of transitions is determined by the energy difference between Landau levels and is therefore controlled by the magnetic and electric fields. The other determining factor is the strength of the spin-orbit gap.  Thus, a careful tuning of $B$ and $E_z$ should allow for a determination of $\Delta_{\rm so}$.

\begin{figure}[h!]
\begin{center}
\includegraphics[width=1.0\linewidth]{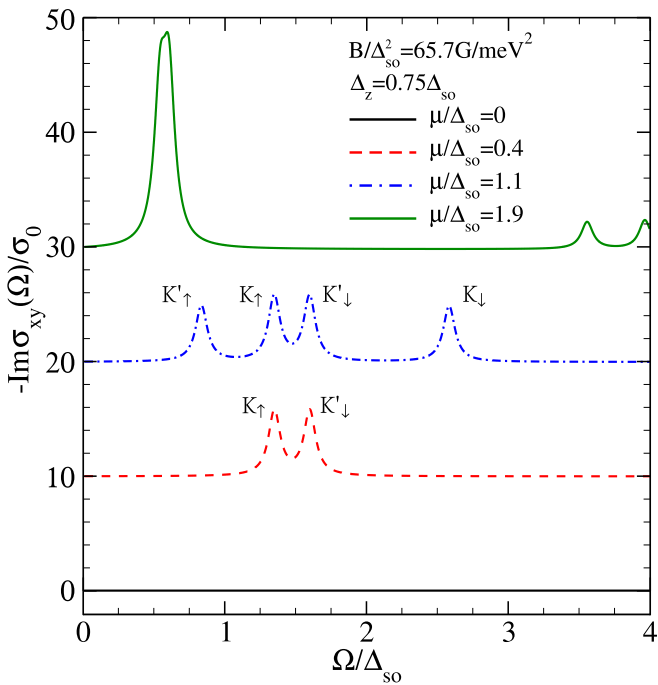}
\end{center}
\caption{\label{fig:Condxy}(Color online) Imaginary part of the zero temperature transverse Hall conductivity of silicene for varying chemical potential for $B/\Delta_{\rm so}^2=65.7$G/meV$^2$, $\eta=0.05\Delta_{\rm so}$ and $\Delta_z=0.75\Delta_{\rm so}$.  These results correspond to those in Fig.~\ref{fig:Cond-mu}(a) and are vertically offset by 10 units.
}
\end{figure}

The absorptive part of the transverse Hall conductivity is found by an evaluation of Eqn.~\eqref{Condxy}.  The results of $\Delta_z=0.75\Delta_{\rm so}$ for varying chemical potential are shown in Fig.~\ref{fig:Condxy}.  
The values shown here correspond to those in Fig.~\ref{fig:Cond-mu}(a) for Re$\sigma_{xx}(\Omega)$.  Im$\sigma_{xy}(\Omega)$ is related to the negative of the longitudinal response. However, not all the features in the longitudinal conductivity have counterparts in the transverse Hall conductivity; indeed, the only features that are present in Im$\sigma_{xy}(\Omega)$ result from transitions between $n$ and $|n|\pm 1$ when the transition from $n\pm 1$ to $|n|$ is Pauli blocked.  Thus, the absorptive part of the transverse Hall conductivity contains a maximum of four features.  For example, when $\mu=0$, Im$\sigma_{xy}(\Omega)$ is zero for all $\Omega$ as no transitions meet the criteria for a finite Hall response.  When $\mu=0.4\Delta_{\rm so}$, the only finite Hall response is due to transitions between the spin up $n=0$ and 1 Landau levels at the $K$ point and $n=0$ and 1 spin down levels at $K^\prime$ as the equal energy transitions from $n=-1$ to 0 spin down levels at $K^\prime$ and $n=-1$ to 0 spin up levels at $K$ are Pauli blocked (refer to Fig.\ref{fig:Trans}).  For $\mu=1.1\Delta_{\rm so}$, four transitions meet the criteria for a finite response and, hence, four features are present in the conductivity.  Identically to Re$\sigma_{xx}(\Omega)$, four valley-spin-polarized absorption lines can be observed.  This can be understood by examining Eqn.~\eqref{Condxy} with particular attention to the minus sign between the two Kronecker $\delta$-function terms.  The negative sign results in a zero contribution when the two prefactors are the same.  The transverse Hall results become important when considering circularly-polarized light.

\section{Results for Circularly-Polarized Light}

\begin{figure}[h!]
\begin{center}
\includegraphics[width=1.0\linewidth]{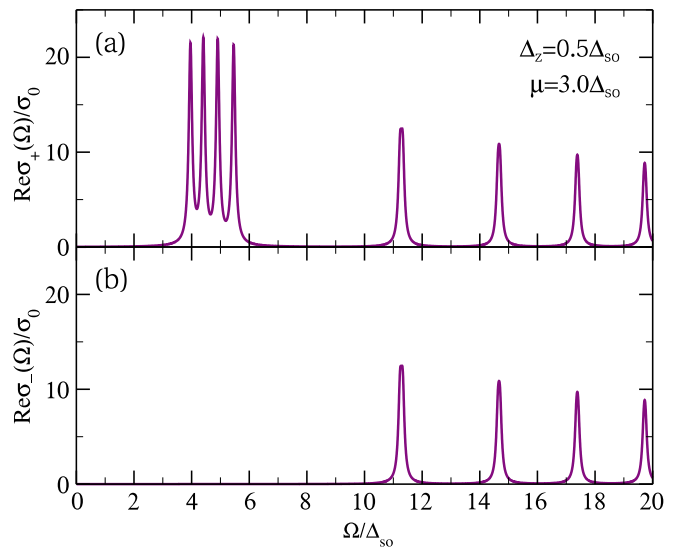}
\end{center}
\caption{\label{fig:Cond-circ}(Color online) The absorptive part of the conductivity response to (a) right-handed and (b) left-handed circularly-polarized light.  In the response to right-handed polarization, the quartet of peaks that meet the criteria for a finite transverse Hall response are of double weight; they are absent in the response to left-handed polarization.  All higher features look identical to the longitudinal response.
}
\end{figure}

In the circularly-polarization basis, the conductivity is given by $\sigma_{xx}(\Omega)\pm i\sigma_{xy}(\Omega)$ for right handed ($+$) and left handed ($-$) polarization\cite{Pound:2012}.  The absorptive part of the conductivity is therefore
\begin{equation}\label{circ}
\text{Re}\sigma_{\pm}(\Omega)=\text{Re}\sigma_{xx}(\Omega)\mp \text{Im}\sigma_{xy}(\Omega).
\end{equation} 
This is easily evaluated with Eqns.~\eqref{Condxx} and \eqref{Condxy}.  The conductivities for right- and left-handed polarization are shown in Fig.~\ref{fig:Cond-circ}(a) and (b), respectively.  
Here, $\Delta_z/\Delta_{\rm so}=0.5$, $B/\Delta_{\rm so}^2=657$G/meV$^2$ and $\mu/\Delta_{\rm so}=3.0$ such that the Fermi energy lies between the $n=0$ and 1 Landau levels.  These values yield the four aforementioned spin- and valley-polarized responses.  A larger magnetic field has been used for a clearer separation of features.  As it is only the quartet of polarized peaks that meets the criteria for a nonzero transverse Hall response, they double in spectral weight for the right-handed response and are not present for left-handed polarization.  In all cases, the higher-energy features look identical to those of the longitudinal response\cite{Tabert:2013a}.

\section{Semiclassical Cyclotron Resonance}

The semiclassical limit occurs when the Landau level spacing becomes inconsequential\cite{Onsager:1952, Pound:2012}.  This occurs for a large chemical potential, $\mu\gg\varepsilon_1$.  Consider $\mu$ between the $N^{th}$ and $N+1^{th}$ Landau levels for $N\gg 1$.  Due to the $\sqrt{n}$ spacing of levels, the frequency of the intraband transition, given by $\delta \varepsilon\equiv \varepsilon_{N+1}-\varepsilon_N$, can be approximated as
\begin{equation}
\delta\varepsilon\approx\frac{v^2\hbar e B/c}{\sqrt{\Delta_{\xi\sigma}^2+2Nv^2\hbar e B/c}}. 
\end{equation}  Noting that $\mu\approx\varepsilon_N$, we obtain
\begin{equation}\label{Cyclotron}
\delta\varepsilon\approx \frac{v^2\hbar e B}{\mu c}\equiv\omega_{\rm cr},
\end{equation}
which is the semiclassical cyclotron resonance frequency\cite{Onsager:1952, Novoselov:2005, Zhang:2005, Pound:2012}.

\begin{figure}[h!]
\begin{center}
\includegraphics[width=1.0\linewidth]{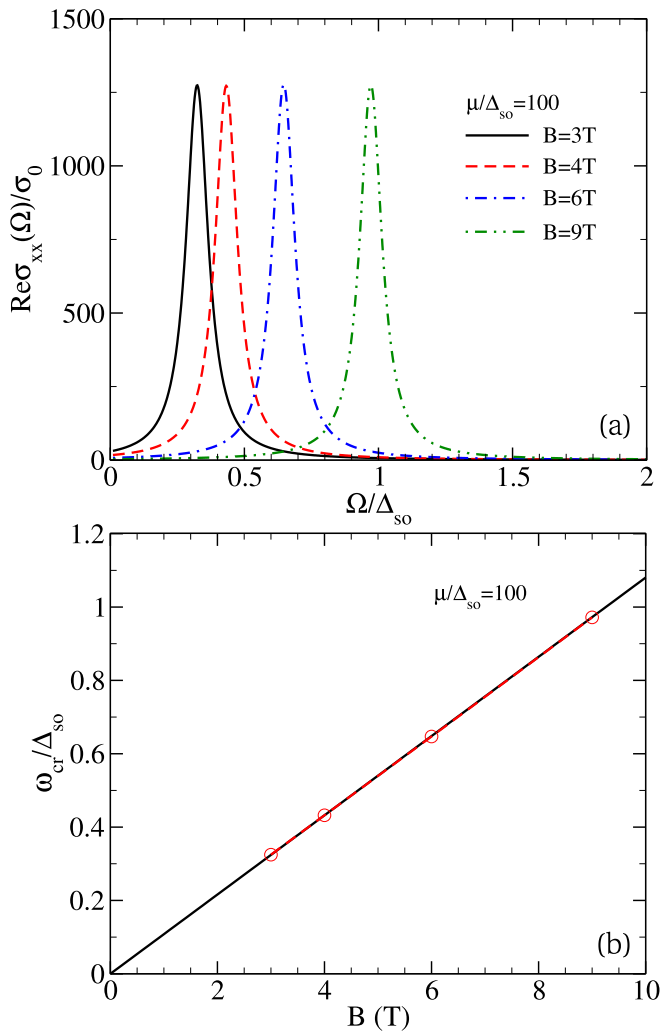}
\end{center}
\caption{\label{fig:Cyclotron}(Color online) (a) The onset of the strong intraband absorption peak for various magnetic fields. (b) A comparison of the actual cyclotron resonance frequency (red circles) found by the onset frequencies in (a) with Eqn.~\eqref{Cyclotron} (solid black line).  The parameters used are $\mu/\Delta_{\rm so}=100$, $\Delta_z=0$ and $\eta/\Delta_{\rm so}=0.05$.
}
\end{figure} 
The cyclotron resonance frequency is also found by examining the onset frequency of the strong intraband absorption peak for large chemical potential.  Several representative examples are shown in Fig.~\ref{fig:Cyclotron}(a).  In Fig.~\ref{fig:Cyclotron}(b), the location of these peaks is compared to the linear formula given by Eqn.~\eqref{Cyclotron}; excellent agreement is found.  The parameters used are $\mu/\Delta_{\rm so}=100$ and $\Delta_z=0$.  While the results are general for all $\Delta_{\rm so}$, a value of 3.9 meV was used so that the magnetic fields in Fig.~\ref{fig:Cyclotron}(a) can be quoted in Tesla.  As this is only valid in the high $\mu$ regime, the value of $\Delta_z$ becomes inconsequential and the graphene results of Ref.~\cite{Novoselov:2005, Zhang:2005,Pound:2012} are obtained.

\section{Conclusions}

Using an effective low-energy Hamiltonian, the magneto-optical conductivity of silicene is computed.  The effects of varying the chemical potential and perpendicular electric field are examined. Particular attention is given to the different insulating phases of the system (topological or band insulator).  Indeed, it is found that the band gap behaviour of the $B=0$ system is a determining factor in the onset and behaviour of transitions in the magneto-optics.  That is, in the topological insulator regime, the strong interband signatures of the $\Delta_z=0$ case split into two with both moving in opposite directions for increased electric field.  As the system transitions into the BI regime, all features move higher in energy.  These signatures should allow for a confirmation and identification of the two insulating regimes as well as provide a measure for $\Delta_{\rm so}$.  With varying chemical potential, a strong spectral weight redistribution is observed with the onset of strong intraband transitions.  By tuning the chemical potential for finite $\Delta_z$, four spin- and valley-polarized responses can be obtained\cite{Tabert:2013a}.   Similar results are found in the response to circularly-polarized light; however, the polarized quartet of absorption peaks is only present in the response to one type of circular polarization.  The onset energies of all features can be tuned by both the magnetic and electric fields.  The ability to produce spin- and valley-polarized charge carriers is of potential interest to spin- and valleytronic technologies.  The semiclassical cyclotron resonance frequency is also obtained and is shown to be well approximated by a linear dependence on magnetic field.  As silicene samples become more available, we believe this work will help guide the development of experimental literature on 2D crystals with sizable spin-orbit interactions.     

\begin{acknowledgments}
We thank J.P. Carbotte for discussions.  This work has been supported by the Natural Science and Engineering Research Council of Canada.
\end{acknowledgments}

\bibliographystyle{apsrev4-1}
\bibliography{llsi}

\end{document}